\documentclass[11pt,preprintnumbers,aps,amssymb,nofootinbib,amsmath]
{revtex4}
\usepackage{epsfig,epsf}
\usepackage{bm} 
\newcommand{\beq}{\begin{equation}}
\newcommand{\beql}[1]{\begin{equation}\label{#1}}
\newcommand{\eeq}{\end{equation}}
\newcommand{\bea}{\begin{eqnarray}}
\newcommand{\eea}{\end{eqnarray}}
\newcommand{\eq}[1]{(\ref{#1})}
\newcommand{\fig}[1]{Fig.~\ref{#1}}
\renewcommand{\sec}[1]{Sec.~\ref{#1}}
\newcounter{topiccounter}
\renewcommand{\b}[1]{\mathbf{#1}}

\newcommand{\im}{\mathrm{Im}\,}

\begin{document}

\preprint{RBRC-814}

\title{Nonlinear pair production in scattering of photons on ultra-short laser pulses at high energy}

\author{Kirill Tuchin$\,^{a,b}$\\}

\affiliation{
$^a\,$Department of Physics and Astronomy, Iowa State University, Ames, IA 50011\\
$^b\,$RIKEN BNL Research Center, Upton, NY 11973-5000\\}

\date{\today}


\begin{abstract}
We consider scattering of a photon on a short intense laser pulse at high energy. We argue that for ultra-short laser pulses the interaction is coherent over the entire length of the pulse. At low pulse intensity $I$ the total cross section for electron-positron pair production is proportional to $I$. However, at pulse intensities higher than the characteristic value $I_s$, the total cross section saturates -- it becomes proportional to the logarithm of intensity. This nonlinear effect is due to multi-photon interactions. We derive the total cross section for pair production at high energies by resuming the multi-photon amplitudes to all orders in intensity. We calculate the saturation intensity $I_s$ and show that it is significantly lower than the Schwinger's critical value.  We discuss possible experimental tests.

\end{abstract}

\maketitle

\section{Introduction}\label{sec:intr}

With advent of very intense laser beams, experimental study of the non-linear Quantum Electrodynamics in photon-photon interactions is becoming feasible. There is a vast literature discussing such non-linear effects in collisions of highly intense beams at low collision energies. Among the expected effects, perhaps the most interesting one is spontaneous electron-positron pair production out of vacuum (Schwinger effect). In the static limit, this effect is predicted to happen at 
intensities of the order of $I_\mathrm{c}=m^4 c^6/(4\pi \alpha \hbar^2)= 10^{29}\,\mathrm{W/cm}^2$ \cite{Sauter:1931zz,Heisenberg:1935qt,Schwinger:1951nm}. At lower intensities the pair-production probability is exponentially suppressed as $e^{-\sqrt{I_c/I}}$, which precludes its experimental observation. If one superimposes low and high frequency fields, then the threshold for the non-perturbative pair production can be lower
than $I_c$ \cite{Schutzhold:2008pz,Monin:2009aj}.

Unlike the non-perturbative mechanism, non-linear pair production in perturbation theory is suppressed only by a power of the ratio $I/I_c$ and may be accessible for observation in the near future. For example, a 
possibility to observe  the non-linear Bethe-Heitler pair production 
was recently discussed in \cite{DiPiazza:2009py,Muller:2008ys}.
One can study this process by colliding an energetic proton or nucleus with photon. This process is possible if the center-of-mass energy $\surd s_{p\gamma}$ is higher than the pair production threshold $2m$.  At higher energies, the virtual photon emitted by the  proton can be thought of as  the equivalent flux of quasi-real photons with spectrum $dn(\hbar\omega/\varepsilon_p)$, where $\hbar\omega$ and $\varepsilon_p$ are  the equivalent photon and proton energy correspondingly. The total cross section of the Bethe-Heitler pair-production is then a convolution of the total photon-photon cross section and the equivalent photon spectrum. 
The former is usually calculated under implicit assumption that  the center-of-mass energy $\surd s_{\gamma\gamma}$  of photon-photon scattering is not too much higher than the threshold $2m$. 

In this paper we consider non-linear pair production in photon-photon  collisions at high energies $\surd s_{\gamma\gamma}\gg 4m$. Unlike,  the  low energy limit, where   one-loop diagrams  give the leading contribution, at high energies these diagrams are suppressed\footnote{Some processes in very intense laser beams may be sensitive to the higher order contributions even at low energies \cite{Kharzeev:2006wg}.}. Indeed, in the leading order in the perturbation theory,  photon-photon scattering is an inelastic process in which photons annihilate into the  electron--positron pair \cite{Breit:1934}. Since this process is mediated by a fermion, the total cross section at high energies decreases with energy as $\sigma_\mathrm{tot}\sim \alpha^2/s$. To be sure, the total cross section of a process mediated by a particle of spin $j$  scales with energy as $\sigma_\mathrm{tot}\sim s^{2j-2}$.
Therefore, in order to find a high energy asymptotic we need to consider a process mediated by photons, even though this process is of higher order in $\alpha$. In such a process incoming photons split into electron-positron $e^-e^+$  pairs that interact via the photon exchange, see \fig{fig:d-d}.  The total cross section for this process is of  order $\alpha^4$; it is energy independent and is given by  Eq.~\eq{gg-fino}  \cite{Lipatov:1969sk,Cheng:1970ef}.

\begin{figure}[ht]
      \includegraphics{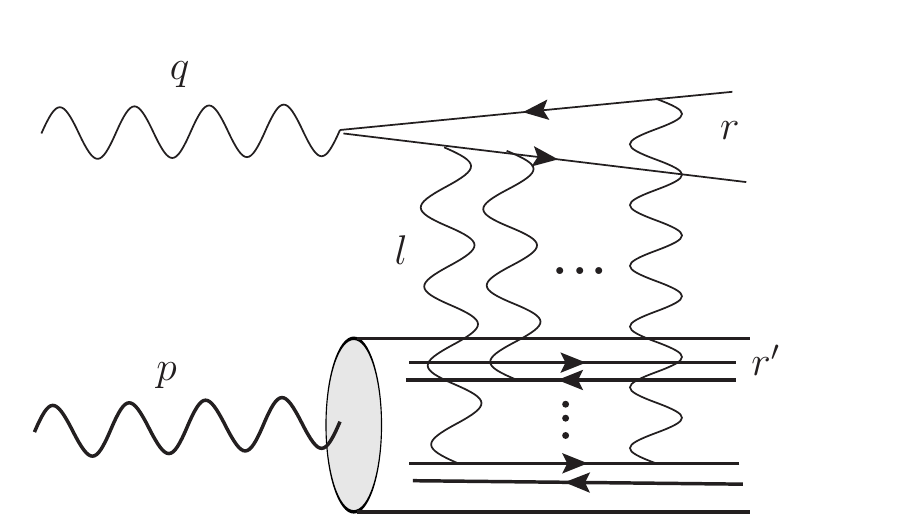} 
  \caption{ Interaction of an energetic photon (upper wavy line) with an ultra-short intense laser pulse (lower wavy line). Horizontal ellipses indicates summation over exchanged $t$-channel photons, vertical ellipses indicate summation over all photons in the laser pulse.  Note that the $t$-channel photons can hook up to either fermion or anti-fermion lines (only one configuration is shown).}
\label{fig:multi}
\end{figure}

An important aspect  of the nonlinear Bethe-Heitler pair production off nuclear targets is strong coherence effect. At the classical level this leads to a characteristic $Z^2$ dependence of the total cross section, where $Z$ is the number of protons in the nucleus. Quantum  coherence effects produce a more complicated $Z$-dependence which can be interpreted as  a result of multiple scattering of  $e^-e^+$ pair in the nucleus \cite{Bethe:1954zz,Ivanov:1998dv,Ivanov:1998ru,Baltz:2001dp,Baltz:2006mz,Lee:2001ea,Bartos:2001jz,Baur:2007zz,Ivanov:1998ka,Tuchin:2009sg}. In this paper we study a similar coherence effect in interaction of a very \emph{energetic photon} $\gamma$ with an  intense \emph{laser pulse} $\ell$. As I recently demonstrated \cite{Tuchin:2009ir},  multiple scattering of $e^-e^+$ pair  in a very intense laser pulse produces  strong non-linear effect on the pair production cross section. The corresponding multi-photon diagram is depicted in \fig{fig:multi}.
The main goal of this paper is to present a detailed derivation and study the properties of  the total cross section for $\gamma\ell$ interaction at high energies including the multiple scattering effect to all orders in the perturbation theory.\footnote{Coherence effects in the fermion mediated pair production in intense laser beams are extensively discussed in \cite{Ritus-dissertation}.}

The paper is structured as follows. In \sec{sec:geom} we introduce the  coherence length $l_c$ which is a distance along the energetic photon propagation direction over which pair production is coherent. When $l_c$ is much larger than the laser pulse length, all $N$ laser pulse photons interaction coherently. Such interactions are not necessarily suppressed in the perturbation theory, because the fine structure constant $\alpha$ is always accompanied by large photon occupation number $N$, which is proportional to the beam intensity; the corresponding expansion parameter $\kappa$ is given by \eq{kk2} and \eq{kapp1}. Thus, a consequence of coherence is that at high pulse intensities multiple scattering of the energetic photon on laser pulse photons becomes an important effect. Because of multiple scatterings the pair-production cross section becomes weakly --  logarithmically -- dependent on the laser pulse intensity. Transition from the  linear to the non-linear regime is characterized  by the \emph{saturation intensity} $I_s$ given by Eq.~\eq{int-cr}, which is inversely proportional to duration and wavelength of the laser pulse.  We will show that for a reasonable choice of parameters $I_s\ll I_\mathrm{c}$. 
The role of the collision geometry in photon-photon interactions was previously discussed in  \cite{Ritus-dissertation,Baier:1989ar}.

In \sec{gg} we introduce the impact parameter representation and show how it can be used to derive the pair--production cross section in the linear (Born) regime.  The advantage of this representation is that the multi-photon interactions take rather simple form in it. They 
are taken care of in \sec{glauber} where we employ the Glauber-Gribov model of multiple scatterings \cite{Glauber:1987bb,Gribov:1968jf,Gribov:1968gs} and treat the laser pulse as a  perfectly  coherent state (in the quantum optics sense).  The main results are presented in \sec{resum} were we derive the total pair--production cross section that  includes any number of multi-photon interactions  given by Eq.~\eq{bsd} and the saturation intensity $I_s$ given by   Eq.~\eq{int-cr}. In \sec{exp} we  discuss the feasibility of experimental study.
Finally, we summarize our results in \sec{summary}.

\section{Geometry of the collision}\label{sec:geom}

\begin{figure}[ht]
      \includegraphics[height=5cm]{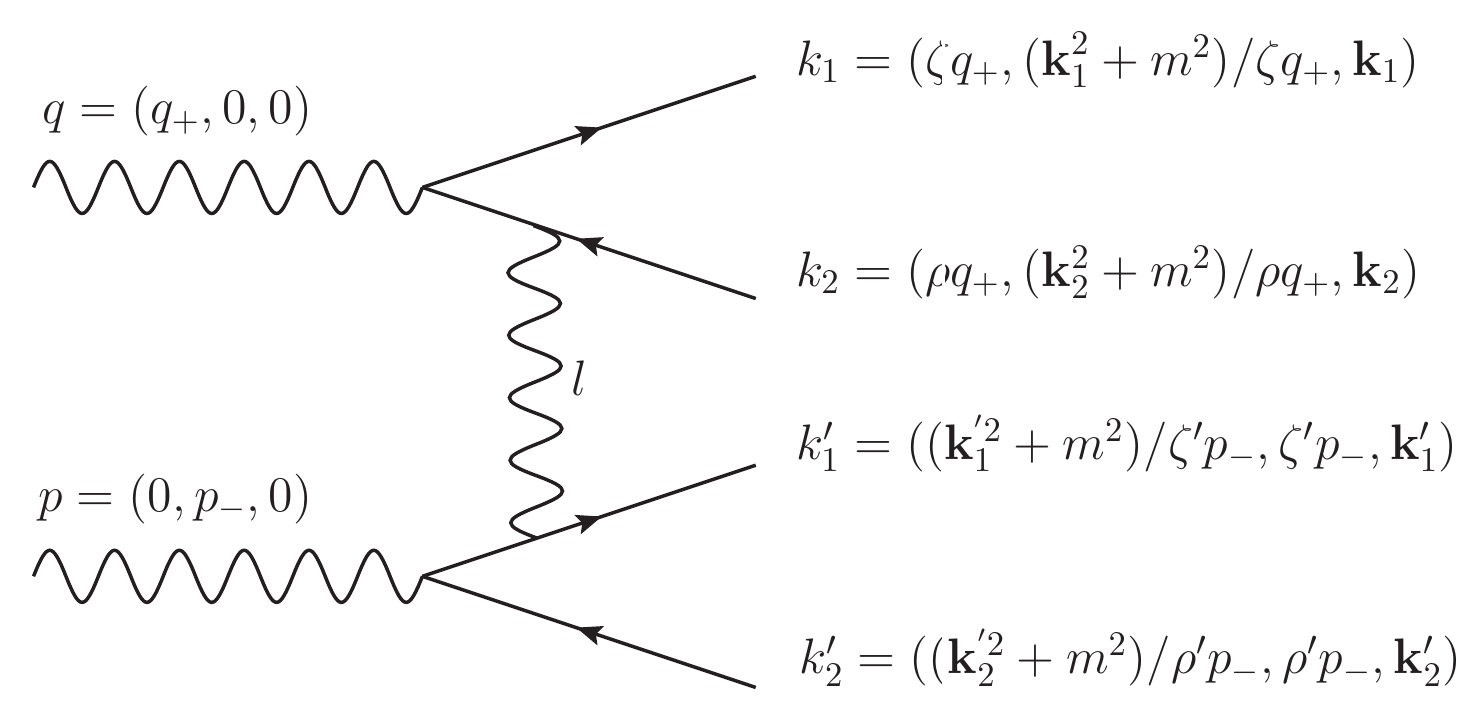} 
  \caption{Pair production in photon-photon scattering. Four-momentum components are  $P=(P_+,P_-,\b P)$, where $\b P$ is the transverse component.}
\label{gg2eeee}
\end{figure}

Let the energetic photon propagate along the ``$-$" light-cone direction with large $q_+$ momentum  ($q_\pm=q_0\pm q_z$, with $z$ being the collision axes), while the laser pulse -- along the ``$+$" light-cone direction with large $p_-$ momentum. Momenta of the produced electrons and positrons are labeled in \fig{gg2eeee}. 
Conservation of the ``$+$" momentum component implies that
\beql{cons}
\zeta q_++\rho q_++\frac{\b k_1^{'2}+m^2}{\zeta' p_-}+\frac{\b k_2^{'2}+m^2}{\rho' p_-}= q_+\,.
\eeq
Dividing both sides by $q_+$ and using $s_{\gamma\gamma}=p_-q_+$ we get in the high energy approximation
\beql{cons1}
\zeta +\rho +\frac{\b k_1^{'2}+m^2}{s_{\gamma\gamma}\zeta' }+\frac{\b k_2^{'2}+m^2}{\rho' s_{\gamma\gamma}}\approx \zeta +\rho=1\,.
\eeq
Analogously, from the conservation of the ``$-$" momentum component it follows that $\zeta' +\rho'=1$. 
Momentum transfer $l_-$ in the  ``$-$" light-cone direction  is given by 
\beql{momtr1}
l_-= \frac{\b k_1^{2}+m^2}{\zeta q_+}+\frac{\b k_1^{2}+m^2}{(1-\zeta) q_+}+\zeta'p_-+(1-\zeta')p_- - p_-= \frac{m^2+\b k_1^2(1-\zeta)+\b k_2^2\zeta}{\zeta(1-\zeta)q_+ }
\eeq
The typical value of the electron and positron transverse momenta is $|\b k_1|\sim  |\b k_2|\sim m$ implying that 
\beq\label{x1}
l_-\simeq \frac{2m^2}{\zeta(1-\zeta)q_+ }\,.
\eeq

Photon-photon scattering amplitude is proportional to the phase factor $\exp\{i\,l\cdot x_a\}$, where $l$ is the four-momentum transfer  and $x_a$ denotes the four-position vector of the interacting particles. The total cross section involves integrals of the type (see e.g.\ \eq{gmp})
\beql{vst1}
\int_0^1d\zeta \, e^{i\,\frac{1}{2}(x_{a+}-x_{b+})\,l_-}=\frac{1}{2}\int_0^1\frac{dy}{\sqrt{1-y}} \, e^{i\,\frac{4\theta}{y}}\,\to \bigg\{ 
\begin{array}{lc}
\frac{\sqrt{\pi }}{4\sqrt{\theta}}e^{-i\theta-\frac{i\pi}{4} }\,, & \theta\gg 1\\
1\,, & \theta\ll 1
\end{array}
\eeq
where  $y=4\zeta(1-\zeta)$ for $0\le \zeta\le 1$  and we denoted $\theta= (x_{a+}-x_{b+})m^2/q_+$. We can see that for small $\theta$ interaction is coherent over the entire laser pulse.  Hence we define the coherence length (in the ``$+$" light-cone direction) as \cite{Berestetsky:1982aq}\footnote{Not to be confused with the laser beam coherence length!}
\beql{vst2}
l_c= \frac{\omega}{m^2}\,,
\eeq
where we used $q_+=2\omega$.
A characteristic feature of  $l_c$ is that it increases with energy.  Because the laser pulse propagates in the ``$+$" light-cone direction, all laser photons within the distance $\Delta x_+= l_c$ interfere during the interaction with the energetic photon.  The interaction is completely coherent when $l_c$ is much larger than the laser pulse length $L$. One can of course also define the coherence length in the  ``$-$" light-cone direction which is given by $1/l_+$. This quantity however is of no interest since we assume that the energetic photon is a structureless particle. 

It is instructive to plug in some numbers into \eq{vst2} to determine the characteristic energies of the energetic photon and the laser pulse. In usual units,  $l_c= (\hbar\omega/mc^2)\lambdabar$, where $\lambdabar = \hbar/mc= 3.8\cdot 10^{-4}$~nm is the Compton wavelength of electron. Condition for coherence takes form
\beql{maxL}
\frac{\hbar\omega}{mc^2}\gg \frac{L}{\lambdabar}\,.
\eeq
Consider for example a hundred attosecond pulse $\tau= 10^{-16}$~sec which translates into $L= \tau c= 30$~nm. Eq.~\eq{maxL} is satisfied for $\hbar\omega\gg 40$~GeV. For a typical attosecond pulse with frequency in the xuv domain \ $\hbar\Omega\simeq 100$~eV this condition also guarantees that 
the high energy approximation holds. For instance,  the center-of-mass energy for $\hbar\omega=200$~GeV is  $\sqrt{s_{\gamma\gamma}}= 2\sqrt{\Omega\omega}\approx 9$~MeV, which is sufficiently large compared to $4m=2$~MeV.

Concerning the typical distances in the transverse direction, they are of the order of the electron Compton wavelength $\lambdabar$.
Thus, the two relevant geometric characteristics of the laser pulse are its  longitudinal extent $L$ and its width $2d$. 
The former is determined by the laser pulse duration $\tau$, the later is of the order of the laser beam wavelength $\lambda$. For example, laser pulse frequency  $\hbar\Omega\simeq 100$~eV corresponds to 
    $d\sim \lambda=13$~nm, which is much larger than $\lambdabar$.  
 Therefore, we can treat interaction of an energetic photon with an attosecond laser pulse  as a problem of propagation of an energetic  photon  through the laser pulse ``medium" of length $L$, width $2d$ and density $\rho= N/(\pi d^2 L)$.

\section{Photon-photon scattering at high energies}\label{gg}

It is instructive to consider first $\gamma\gamma$ scattering at high energies but low intensities. Low intensity ensures that the multi-photon processes are suppressed.
\begin{figure}[ht]
      \includegraphics[height=3.5cm]{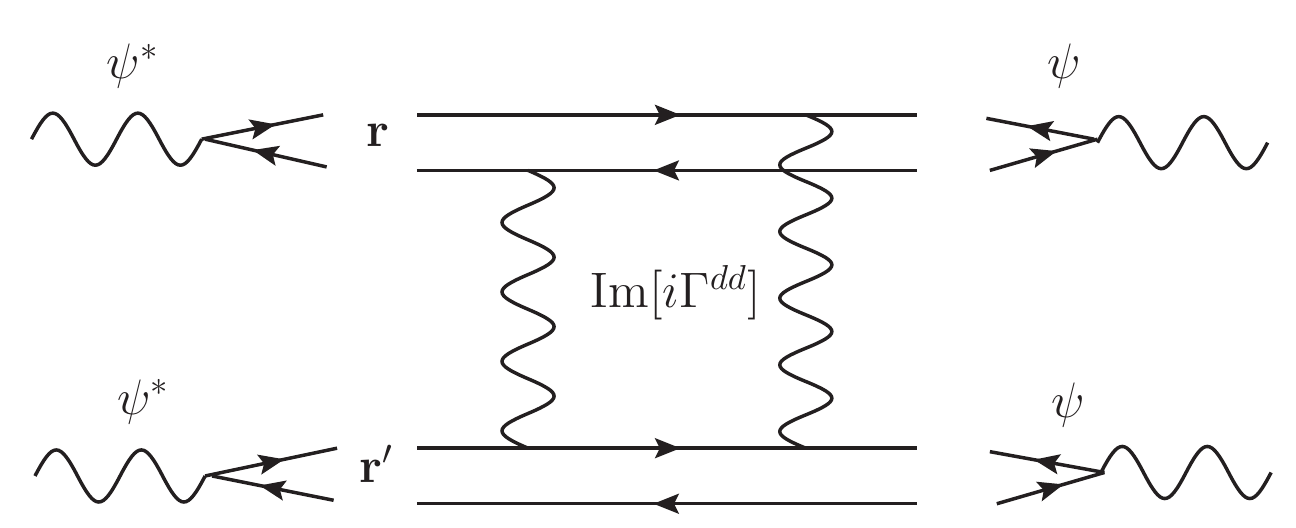} 
  \caption{One of the diagrams contributing to the photon--photon scattering at high energies and low intensities. Other diagrams are obtained by all possible permutations of the $t$-channel photons. $\Phi= |\psi|^2$ with $\Phi$ given by \eq{phi}.    }
\label{fig:d-d}
\end{figure}
The total $\gamma\gamma$ cross section can be written as a convolution of two photon light-cone wave-functions $\Phi$ and the imaginary part of elastic scattering amplitude of two electric dipoles  $\im [i\Gamma^{dd}]$ \cite{Mueller:1994gb,Donnachie:1999kp,Bondarenko:2003yma}, see \fig{fig:d-d}. Photon ``wave function" describes splitting of a photon into a lepton-antilepton pair of transverse size $\b r$ with lepton carrying fraction $\zeta$ of the photon's light-cone momentum. It is given by \cite{Nikolaev:1990ja}
\beql{phi}
\Phi(\b r, \zeta)= \frac{2\,\alpha\, m^2}{\pi}\left\{ K_1^2(rm)[\zeta^2+(1-\zeta)^2]+K_0^2(rm)\right\}\,.
\eeq
Imaginary part of the elastic scattering amplitude of two electric dipoles of sizes $\b r$ and $\b r'$ at impact parameter $\b B$  at the leading order in the perturbation theory reads
\beql{sc.ampl}
\im[i\Gamma^{dd}( \b r, \b r',\b B,s)]= 2 \alpha^2\ln^2\frac{|\b B+\frac{1}{2}\b r+\frac{1}{2}\b r'||\b B-\frac{1}{2}\b r-\frac{1}{2}\b r'|}{|\b B+\frac{1}{2}\b r-\frac{1}{2}\b r'||\b B-\frac{1}{2}\b r+\frac{1}{2}\b r' |} \,.
\eeq
It corresponds  to the two photon exchange, see \fig{fig:d-d}. The total cross section is given by 
\beql{gmp}
\sigma^{\gamma \gamma}_\mathrm{tot}(s)=\frac{1}{2!}\int d^2B\int \frac{d^2r}{2\pi}\int_0^1 d\zeta \,\frac{1}{2}\Phi(\b r,\zeta)\int \frac{d^2r'}{2\pi}\int_0^1 d\zeta' \,\frac{1}{2}\,\Phi(\b r',\zeta') \, 2\,\im [i\Gamma^{dd}( \b r, \b r',\b B,s)]\,,
\eeq
where the symmetry factor $1/2!$ is inserted to avoid double-counting of identical dipoles in the final state of the process $\gamma\gamma\to dd$, with $d=e^-e^+$.
Since the scattering amplitude \eq{sc.ampl} does not depend on $\zeta$ and $\zeta'$, we can integrate over them.\footnote{Had the integral over $\b B$ in \eq{gammap} required a cutoff, $l_c$ would have entered the final result \eq{gg-fino} -- as it does in the Bethe-Heitler pair production -- and then $\zeta$-integration would have been more complicated. However, no such subtlety arises here. }
 It is convenient to introduce an auxiliary function $F$
\beql{faux}
F(x)= \frac{2}{3}K_1^2(x)+K_0^2(x)\,
\eeq
and cast the cross section into the form
\beq\label{gammap}
\sigma^{\gamma \gamma}_\mathrm{tot}(s)=\frac{1}{2!}\left(\frac{2\alpha m^2}{\pi}\right)^2\int d^2B\int \frac{d^2r}{2\pi}\frac{1}{2}F(mr)\int \frac{d^2r'}{2\pi}\frac{1}{2}F(mr')\,2\,\im [i\Gamma^{dd}(\b r, \b r',\b B,s)]\,.
\eeq
These integrals are performed in Appendix~\ref{appA} with the result:
\beql{gg-fino}
\sigma_\mathrm{tot}^{\gamma\gamma}(s)= \frac{\alpha^4}{ m^2}\frac{1}{36\pi}[175\zeta(3)-38]\,.
\eeq
We observe that the cross section is energy independent. Thus, we  re-derived the well-known result for the total cross section \cite{Lipatov:1969sk,Cheng:1970ef} using the impact parameter representation. 


\section{Glauber-Gribov model for multi-photon interactions}\label{glauber}

As noted in Introduction, from the point of view of a high energy photon interacting with the ultra-short laser pulse, the later can be considered as a  neutral medium characterized by a certain number density $\rho$. Scattering off such a medium involves multiple simultaneous interactions with many laser pulse photons. A general approach to this kind of problems was developed by Glauber \cite{Glauber:1987bb} and Gribov \cite{Gribov:1968jf,Gribov:1968gs} and we adopt this theory for the case in hand. 

\begin{figure}[ht]
      \includegraphics[width=3cm]{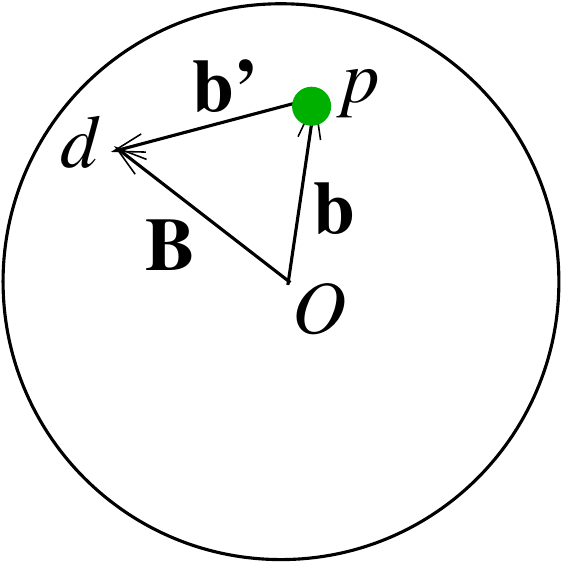} 
  \caption{Geometry of the photon-laser pulse interaction in the  configuration space. $d$ is the dipole associated with the energetic photon, $p$ is a photon in the laser pulse, $O$ is the position of the symmetry axes (orthogonal to the page). }
\label{fig:geom}
\end{figure}

Elastic scattering amplitude of a dipole -- produced by the energetic photon -- on a laser pulse  $i\Gamma^{d\ell}$ is simply related to the scattering matrix element $S$ as $\Gamma^{d\ell}(s,\b B)= 1-S^{d\ell}(s,\b B)$. Laser pulse is a coherent state, therefore  interactions of the dipole $d$  with different  photons in the laser pulse are independent.
This implies that  only two-body interactions must be taken into account, while neglecting the many-body forces \cite{Glauber:1987bb}. Thus, the laser pulse average of the amplitude reads
\beql{av.nuc}
\langle \Gamma^{d\ell}(\b r, \b B , s)\rangle = \langle 1- S^{d\ell}(\b r, \b B , s)\rangle 
= 1- \langle S^{d\gamma_\ell}(\b r, \b B , s)\rangle ^N=
1- [1-\langle \Gamma^{d\gamma_\ell}(\b r, \b B , s)\rangle ]^N\,,
\eeq
The average in \eq{av.nuc} is performed using the number density $\rho(\b b,z)$ of photons $\gamma_\ell$ in the pulse  normalized such that 
\beq\label{dens.norm}
\int d^2b\, dz\,\rho(\b b,z)= N\,.
\eeq
We have
\beql{a1}
\langle i \Gamma^{d \gamma_\ell}(\b r, \b B, s)\rangle= \frac{1}{N}\int d^2b\int_{-L/2}^{L/2} dz\, \rho\,i \Gamma^{d \gamma_\ell}(\b r, \b b', s)
\,,
\eeq
where $\b b'$, $\b B$ and $\b b$ are  the impact parameters between the dipole $d$ and photon $\gamma_\ell$,  dipole $d$ and the laser  pulse symmetry axes and photon $\gamma_\ell$ and the symmetry axes, see \fig{fig:geom}. Obviously, $\b b'=\b B-\b b$.
Assuming for notational simplicity  that $\rho$ is independent of $z$ we get
\beql{a2}
\langle i \Gamma^{d\gamma_\ell}(\b r, \b B, s)\rangle= \frac{L}{N}\int d^2b\, \rho(\b b)\,
i \Gamma^{d \gamma_\ell}(\b r, \b B-\b b, s)
\,.
\eeq

The limit that we are interested in is $N\gg 1$ while $N\langle \Gamma^{d\gamma_\ell}\rangle\sim 1$. Taking this limit in \eq{av.nuc} produces 
\beql{aZ}
\langle \Gamma^{d\ell}(\b r, \b B , s)\rangle =1- \exp\{-N\langle \Gamma^{d\gamma_\ell}(\b r,\b B,s)\rangle\}
\,.
\eeq
The problem of calculating the scattering amplitude of the dipole $d$ on a system of $N$ laser photons thus reduces to the problem of calculating the scattering amplitude of the dipole $d$ on a single photon $\Gamma^{d\gamma_\ell}$. 

We estimated in \sec{sec:geom} that $b'\ll d$. Because $b \sim d$ it implies that $b'\ll B\approx b$. Changing the integration variable in \eq{a2} $\b b\to \b b'$ we derive
\beql{a3}
\langle i \Gamma^{d\gamma_\ell}(\b r, \b B, s)\rangle\approx \frac{L}{N}\,\rho(B) \int d^2b' i \Gamma^{d \gamma_\ell}(\b r,\b b', s)\,.
\eeq
Employing the dipole representation with respect to the laser photon $\gamma_\ell$ we write
\beql{tt}
i\Gamma^{d \gamma_\ell}(\b r,\b b', s)= 
\int \frac{d^2r'}{2\pi}\int_0^1d\zeta' \,\frac{1}{2} \Phi(r',\zeta')\,i \Gamma^{dd}(\b r,\b r',\b b',s)\,.
\eeq
\begin{figure}[ht]
      \includegraphics[width=7cm]{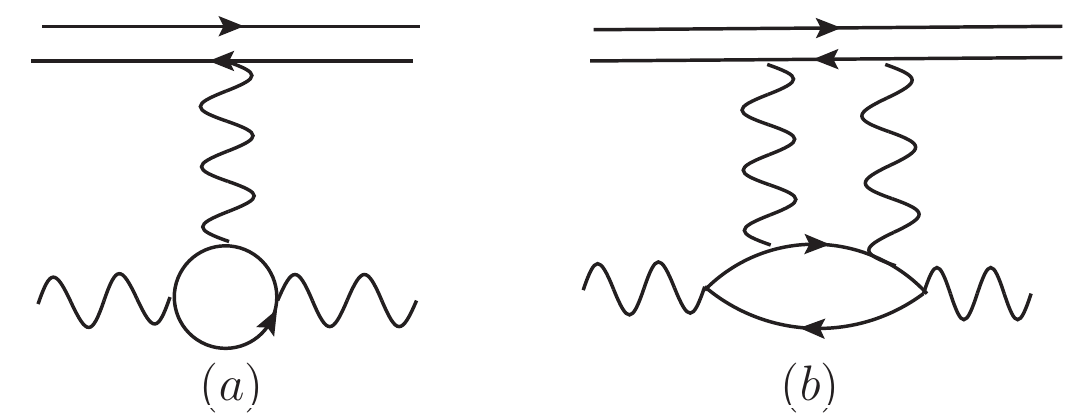} 
  \caption{Leading order contributions to the dipole-photon \emph{elastic} cross section. The upper dipole is associated with the energetic photon. The horizontal wavy line is  a photon in the  laser pulse.}
\label{fig:gamma}
\end{figure}
The $\b b'$-integral in \eq{a3} can be taken using \eq{phases2}. 

Diagram \fig{fig:gamma}(a) represents the leading contribution to the real part of the dipole-photon scattering amplitude. It vanishes due to the $C$-invariance of QED (Furry's theorem). Higher order contributions to the real part of the dipole-photon amplitude are given by diagrams with odd number of exchanged photons. They vanish for the same reason. Therefore, the  dipole-photon amplitude is purely imaginary at high energies. The leading diagram is shown in \fig{fig:gamma}(b). The corresponding dipole-dipole amplitude is given by \eq{sc.ampl}.

\section{Saturation at high intensities}\label{resum}

Since at high energies coherence length $l_c$ becomes larger than the laser pulse length $L$,  the dipole $d$ can simultaneously interact with any number of photons $\gamma_\ell$. To take this into account we use the unitarity relation applied to the elastic scattering amplitude at given impact parameter. The unitarity relation reads  
\beql{unitarity}
2\,\im(i \Gamma^{d\ell})= | \Gamma^{d\ell}|^2+G^{d\ell}\,,
\eeq
where $G^{d\ell}$ stands for inelastic scattering amplitude. Using \eq{aZ} we can solve \eq{unitarity} as
\beql{gin}
G^{d\ell}= 1-e^{-N \langle \im(i\Gamma^{d \gamma_\ell})\rangle } \,.
\eeq
It follows that the total dipole--laser pulse cross section is given by
\beq\label{secs}
\sigma_\mathrm{tot}^{d\ell}= 2\int d^2B \left\{ 1-  e^{ -N \langle \im(i\Gamma^{d \gamma_\ell})\rangle}\right\} 
\eeq
The total cross section for the $\gamma$-$\ell$ scattering then reads (see \eq{gmp})
\beql{gmp2}
\sigma^{\gamma \ell}_\mathrm{tot}(s)=\frac{1}{2}\frac{2\alpha m^2}{\pi}\,\int d^2B\int \frac{d^2r}{2\pi}F(mr)\,\left\{ 1-  e^{ -N \langle \im[i\Gamma^{d \gamma_\ell}(\b r, \b B, s)]\rangle}\right\}
\eeq
Combining together \eq{sc.ampl},\eq{a3},\eq{tt} and \eq{gmp2} yields
\begin{eqnarray}\label{bsd}
\sigma^{\gamma \ell}_\mathrm{tot}(s)&=&\frac{\alpha}{\pi}\int d^2B \int_0^\infty du\, u\, F(u)\nonumber\\
&&\times \bigg\{ 1- \exp\bigg[-\frac{8\alpha^3\rho(B) L}{m^2} u^4\int_0^1 d\xi\, \ln \frac{e}{\xi}\big( \xi^3F(u\xi)+\xi^{-3}F(u\xi^{-1})\big)\bigg]\bigg\}
\,,
\end{eqnarray}
where $\xi$ and $u$ are dimensionless variables defined in Appendix~\ref{appA}. 

\begin{figure}[ht]
      \includegraphics[height=5cm]{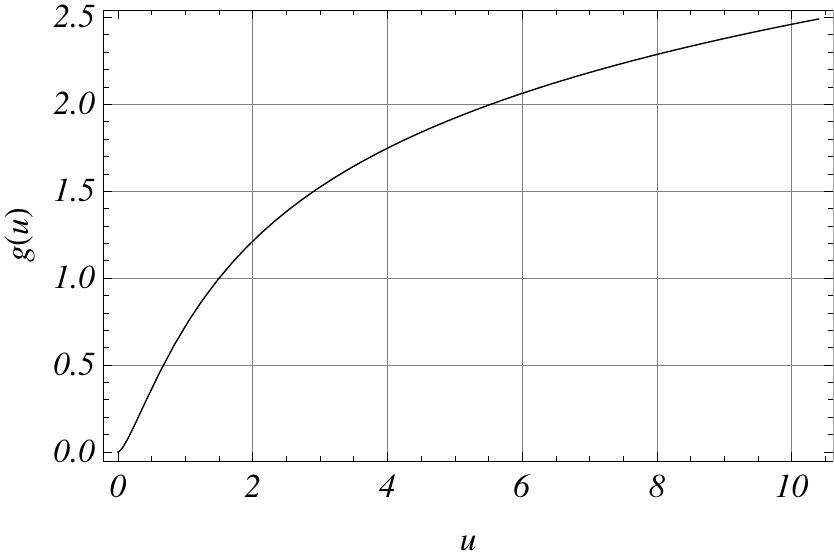}       
  \caption{ Function $g(u)$, see \eq{g}.   }
\label{fig:gu}
\end{figure}
Let us analyze the relative magnitude of the 
the multi-photon processes as compared to the leading Born approximation.
To this end, we introduce parameter 
\beql{kk2}
\kappa= \frac{8\alpha^3\rho L}{m^2}
\eeq
and an auxiliary function
\beql{g}
g(u) = u^4\int_0^1 d\xi\, \ln \frac{e}{\xi}\big( \xi^3F(u\xi)+\xi^{-3}F(u\xi^{-1})\big)
\eeq
and split the integral over $u$ in \eq{bsd} into  two integrals:
 \beq\label{w1}
 \int_0^{u_0}du\,  u\, F(u)\big\{ 1- e^{ -\kappa  g(u)} \big\}
 + \int_{u_0}^\infty du \, u\, F(u)\big\{ 1- e^{ -\kappa  g(u)} \big\}\,,
 \eeq
 where $u_0$ is defined such that  
 \beq\label{unot}
 \kappa \, g(u_0)=1\,.
\eeq
\begin{figure}[ht]
\begin{tabular}{cc}
      \includegraphics[height=5cm]{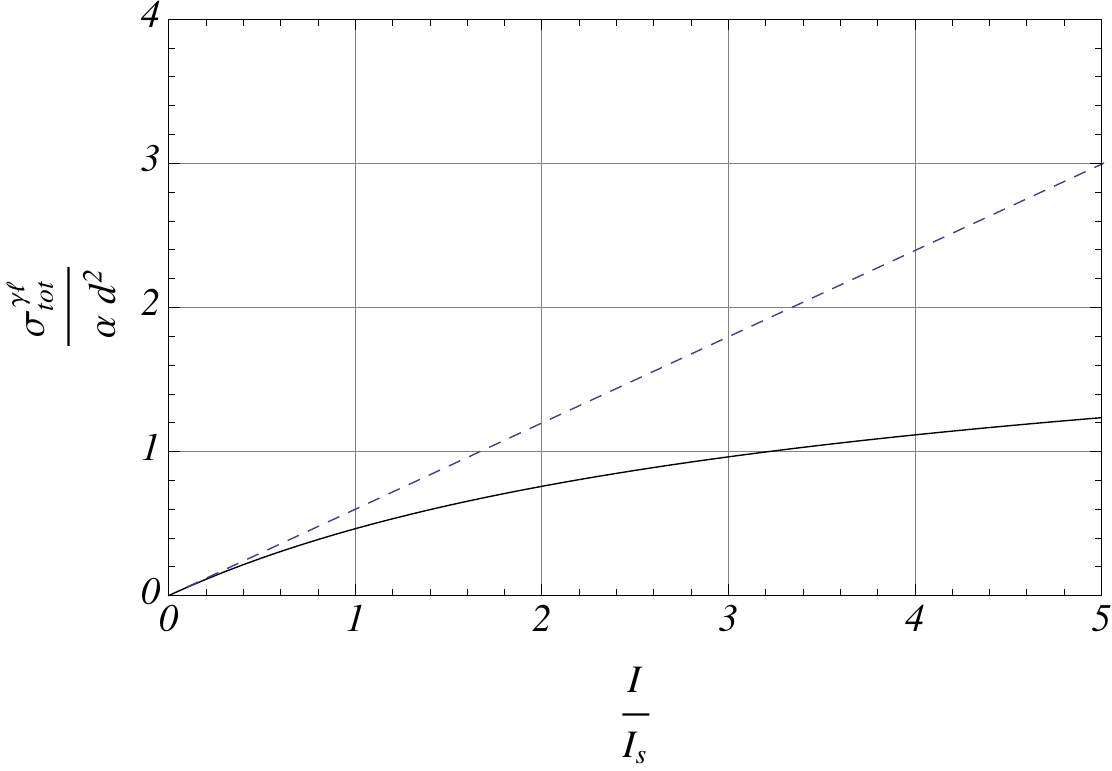} &
      \includegraphics[height=5.4cm]{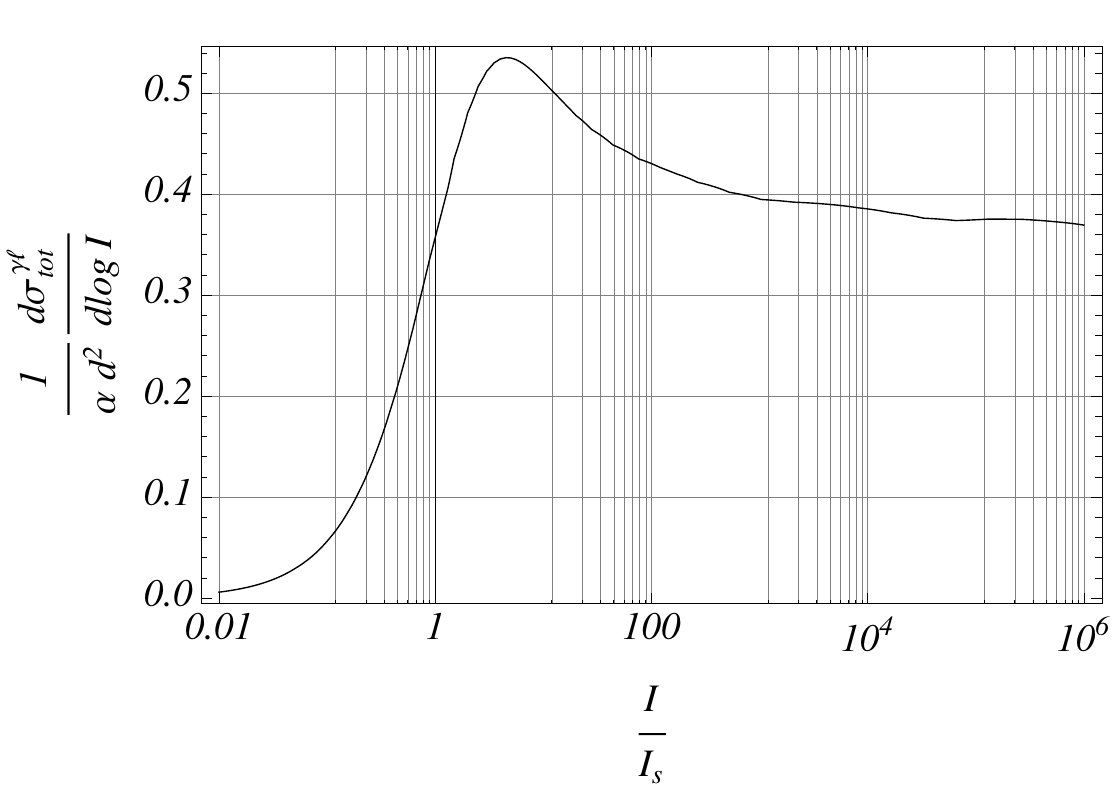}\\
      $(a)$ & $(b)$ 
      \end{tabular}
  \caption{(a) Total $\gamma\ell$ cross section and  (b) its logarithmic derivative as a function of $\kappa= I/I_s$. The broken line in (a) indicates the Born approximation \eq{born}.  }
\label{fig:xsect}
\end{figure}
Let $\kappa\ll 1$, i.e.\ consider a low intensity laser pulse. Function $g(u_0)$ in the l.h.s.\ of Eq.~\eq{unot} is monotonically increasing and $g(1)=0.72$, see \fig{fig:gu}. Moreover, $g(u)\sim \ln u $ at large $u$. Therefore, \eq{unot} can be satisfied only at  $u_0\sim e^{1/\kappa}\gg 1$. On the other hand,  large values of $u$ are suppressed in the integrand of \eq{bsd} because $F(u)\sim e^{-2u}$.  Hence the second term in \eq{w1} can be neglected, while in the first term we can expand the exponent and set the upper limit of integration to infinity. This reproduces the Born approximation (cf.\ \eq{gg-fino})
\beql{born}
\sigma^{\gamma \ell}_\mathrm{tot}(s)= \frac{\alpha^4 N}{ m^2}\frac{1}{36\pi}[175\zeta(3)-38] = \alpha\,\pi d^2\,\kappa\,\frac{1}{288\pi}[175\zeta(3)-38] \,,\qquad \kappa\ll 1\,.
\eeq

Turning to the opposite limit of high intensities $\kappa\gg 1$, we see from Eq.~\eq{unot} that obviously, $u_0\ll 1$  . Then, $g(u)\sim u^2\ln^2 u$ and $u_0\sim 1/\sqrt{\kappa}$. Now,  we can neglect the first term in \eq{w1}, which yields higher orders in $u_0$,  whereas the second term in \eq{w1} gives the following contribution in \eq{bsd}:
\beql{sec.term}
\sigma^{\gamma \ell}_\mathrm{tot}(s)=\frac{\alpha}{\pi}\int d^2B \int_{u_0}^\infty du\, u\, F(u)=\frac{\alpha}{\pi}\,\pi d^2\,\frac{2}{3}\ln\frac{1}{u_0}
=\frac{1}{3}\,d^2\alpha\ln \kappa\,,\qquad \kappa\gg 1
\,.
\eeq
At such high intensities Born approximation breaks down as the multi-photon processes become of the same order of magnitude. 

It is also instructive to look at the logarithmic derivative of the total cross section with respect to $\kappa$. This quantity exhibits a non-trivial pre-asymptotic behavior at large $\kappa$ such that  $\log\kappa\gtrsim 1$. We have using $\kappa\,u_0^2\ln^2u_0\sim 1$ and \eq{bsd}:
\beql{x33}
 \frac{d\sigma^{\gamma \ell}_\mathrm{tot}}{d\ln\kappa}=\frac{1}{2}\,
\frac{d\sigma^{\gamma \ell}_\mathrm{tot}}{d\ln \frac{1}{u_0}}\,\bigg(1-\frac{1}{\ln \frac{1}{u_0}}\bigg)^{-1}\approx  \frac{1}{3}\,\alpha d^2 \, \bigg(1+\frac{1}{\ln \sqrt{\kappa}}\bigg)\,, \qquad \kappa\gg1\,,\quad \ln\kappa\gtrsim 1\,.
\eeq
This behavior is clearly seen in our numerical calculation in \fig{fig:xsect}.

To estimate $\kappa$ defined in \eq{kk2}, note that the average number of photons in the laser pulse of frequency $\Omega$ and total energy  $ E^2 V$ is $N=E^2 V/\Omega$, where $E$ is electric field  and $V$ is volume.  Therefore, the mean photon density is 
\beql{rel-01}
\rho\sim  \frac{E^2}{\Omega}= \frac{I}{\Omega}\,,
\eeq
where $I$ is the laser pulse intensity. Substituting this into \eq{kk2} gives (in usual units)
\beql{kapp1}
\kappa\sim \frac{4 I}{\pi\hbar c}\,\alpha^3\tau \lambdabar^2\lambda\,.
\eeq
Given the values of $L$ (that must be short enough for the formalism of this paper to apply)  and $\Omega$, the value of intensity at which multi-photon effects become important --  \emph{the saturation intensity}  -- reads
\beql{int-cr}
I_s=\frac{\pi \hbar c}{4 \alpha^3\tau \lambdabar^2\lambda}\,.
\eeq
It is convenient to express $I_s$ in terms of the critical  intensity $I_\mathrm{c}$ at which the perturbative QCD vacuum becomes unstable with respect to spontaneous pair-production \cite{Sauter:1931zz,Heisenberg:1935qt,Schwinger:1951nm}. Spontaneous pair-production is another example of the non-linear effect in QED that -- unlike the process considered in this paper --  is intrinsically non-perturbative. From $E_c\sim m^2/e$ we get $I\sim m^4/e^2$. Thus, 
\beql{comp1}
I_s=\frac{\pi^2\,\lambdabar^2}{\alpha^2 \lambda L} \, I_\mathrm{c}\,.
\eeq
For the values of $\lambda$ and $L$ that we use throughout this paper we obtain that $I_s$ is about $10^4-10^5$ times smaller than the value of the critical intensity $I_\mathrm{c}$.
At high laser pulse intensities $I\gg I_s$, the total cross section saturates. The limit is given by Eq.~\eq{sec.term}.

Finally, when $\kappa\gtrsim 1$ contribution of the elastic channel to the total cross section catches up with the inelastic one. Elastic channel corresponds to the processes $\gamma\ell\to e^+e^-\ell$ in which the laser pulse stays intact. The inelastic channel is $\gamma\ell\to e^+e^-e^+e^-$, i.e.\ the laser pulse looses coherence and destroyed. Using Eq.~\eq{unitarity} elastic cross section is given by 
\beq\label{el.x}
\sigma^{\gamma \ell}_\mathrm{el}(s)=\frac{\alpha m^2}{\pi}\int d^2B\int \frac{d^2r}{2\pi}F(mr)\, \big( 1-  e^{ -\frac{1}{2}\kappa g(mr)}\big)^2 \,.
\eeq

\section{Possibility of experimental observation}\label{exp}

As explained in Introduction at asymptotically high energies the process of electro-positron pairs production advocated in this paper is dominant since the corresponding cross section scales as $\sigma^{(1)}\sim s^0$; the superscript $1$ indicates the spin of the mediating particle (photon). This process is characterized by  the multiple pair-production in a single collision $\gamma\ell \to n(e^+e^-),\,\, n\ge 2$.  However, at moderate energies, that are  presently available, it competes with the single pair production $\gamma\ell \to e^+e^-$\footnote{I would like to thank the anonymous referee of this paper for drawing my attention to this point.}, mediated by a fermion. The corresponding cross section at the lowest order is given by 
\beql{sprod}
\sigma^{(1/2)}= \frac{2\pi\alpha^2}{s}\big(\ln \frac{s}{m^2}-1\big)\,.
\eeq 
Comparing this equation with \eq{gg-fino} that gives $\sigma^{(1)}$ at the leading order we conclude that the multi-pair production becomes the leading source of $e^+e^-$ pairs when  $s_{\gamma\gamma}>m^2/\alpha^2$. Let us denote $\beta^2= s_{\gamma\gamma}/(4m)^2=\omega\Omega/(4m^2)$. High energy limit means that $\beta\gg 1$.  Hence we can distinguish the following two kinematic  regions (I) $1\ll \beta\ll 1/\alpha$ where  single pair production dominates and (II) $\beta \gg 1/\alpha$ where  multiple pair-production  discussed in this paper takes over.

For discussion of experimental applications it is  convenient to determine what are the  minimal/maximal values of parameters at  which the process discussed in this paper is possible to observe. For a given laser pulse length $L$, Eq.~\eq{maxL} implies that the minimal energetic photon energy is $\omega_\mathrm{min}= m^2L$. For a given $\beta$ the minimal laser frequency is $\Omega_\mathrm{min}= 4\beta^2/L$.
Then using \eq{comp1} we obtain in usual units
\beq\label{set}
\omega_\mathrm{min}= mc^2\,\frac{L}{\lambdabar}\,,\qquad 
\Omega_\mathrm{min}= \frac{4\beta^2c}{L}\,,\qquad
\bigg(\frac{I_s}{I_\mathrm{c}}\bigg)_\mathrm{min}=\frac{2\pi\,\beta^2}{\alpha^2}\,\bigg( \frac{\lambdabar}{L}\bigg)^2\,.
\eeq 

In the kinematic region (I) one has  to select pairs produced via the multiple pair production mechanism among the majority of pairs produced via the single pair production. We suggest that this can be done by studying the azimuthal correlations of electron and positron. To this end note, that due to momentum conservation, electron and positron produced in the process $\gamma\ell \to e^+e^-$ have momenta $\b k_{e^-}=-\b k_{e^+}$ in the 
center-of-mass frame. On the contrary, the azimuthal correlations in the multiple pair production process are spread over a wider range of azimuthal angles. Thus, by vetoing the pairs correlated back-to-back one performs the required selection. Similar measurements have been done at hadronic colliders. However, assessment  of the practical significance of this method in our case requires numerical simulations  with realistic experimental parameters.

Now, consider for example, 
 a $\tau= 100$~as pulse which corresponds to $L=30$~nm. Then, using \eq{set} we obtain $ \omega_\mathrm{min}= 50$~GeV,   $ \Omega_\mathrm{min}= 200$~eV, where I used $\beta^2=10$. As explained in the Introduction, these parameters are certainly within the experimental reach if one uses the modern portable ultra-short lasers at the  relativistic particle accelerators  \cite{Muller:2008ys}.  Although at present the ratio  $I_s/I_\mathrm{c}\sim 10^{-4}$ is not small enough for such lasers, one may be able to observe small deviations from the linear behavior. 

The kinematic region (II) is ideal for studying the asymptotic behavior of the $\gamma\ell$ scattering. However, for $ \Omega_\mathrm{min}= 200$~eV one needs at least $ \omega_\mathrm{min}\simeq100$~TeV which is beyond the experimental reach. A possible solution would be to employ lasers operating in the $X$-ray region for which a much lower  $ \omega_\mathrm{min}$ will suffice. This option hinges on the availability of portable versions of such devises in future. 
In general,  it would be optimal to have energetic photon with as high energy $\omega$ as possible while the frequency $\Omega$  and pulse duration $\tau$ of the laser pulse as low as possible. Indeed, the lower $\Omega$, i.e.\ the higher $\lambda$, the lower is $I_s$ according to \eq{int-cr}.


\section{Summary}\label{summary}
  
In summary, we studied  pair production  in scattering of energetic photon on an intense  attosecond laser pulse at high energy.  The asymptotic behavior of the corresponding cross section is governed by exchange of virtual photons in the $t$-channel. We resumed all multi-photon amplitudes to derive Eq.~\eq{bsd} which is the main result of our paper. 
The method that we employed  was recently used to obtain some well-known results of QED \cite{Ivanov:1998ka,Tuchin:2009sg} and has been an important tool for studying the high energy QCD for a long time \cite{dipole}.  Nevertheless,  it would be instructive to  re-derive the results of this paper in a more traditional approach based on exact solutions of Dirac equation in an external laser field, see e.g.\  \cite{Ritus-dissertation}. This may give some new insights into the nature of the non-linear pair production at high energies. We plan to  consider this problem elsewhere.

Based on our results we  argued, that the geometry of the  high energy interaction is such that the laser pulse  appears to be wide and short, i.e.\  the energetic photon probes distances in the transverse plane that are much shorter than the pulse width. On the other hand, the interaction coherence length can become larger than pulse length. In this case the energetic photon interacts coherently with the entire laser pulse. At low intensities,  the pair production cross section is proportional to the pulse intensity. However, at intensities $I\gtrsim I_s$ the cross section saturates: it grows only logarithmically with intensity. This behavior is a clear signature of the multi-photon effects.

\acknowledgments
I  am grateful to Genya Levin for many fruitful discussions of related problems. 
This work  was supported in part by the U.S. Department of Energy under Grant No.\ DE-FG02-87ER40371. I 
thank RIKEN, BNL, and the U.S. Department of Energy (Contract No.\ DE-AC02-98CH10886) for providing facilities essential
for the completion of this work.

\appendix
\section{}\label{appA}
Here we evaluate  the integrals appearing in \eq{gammap}. 
Using the identity
\beql{iden}
\int \frac{d^2l}{l^2}\big( e^{i\b a\cdot \b l}-e^{i \b b\cdot \b l}\big)= 2\pi \ln \frac{b}{a}\,,
\eeq
where $a= |\b a|$, we replace the logarithms in the imaginary part of the amplitude by two integrals over momenta $\b l$ and $\b l'$. Integral over  $\b B$ then reduces to the delta function $\delta^{(2)}(\b l-\b l')$  and we obtain 
\begin{eqnarray}
\int d^2B\, \im [i\Gamma^{dd}(\b r, \b r',\b B,s)] 
&=&2\alpha^2 
\int \frac{d^2l}{l^4}\bigg[ e^{i\b l\cdot(\b r-\b r')\frac{1}{2}}-e^{-i\b l\cdot(\b r+\b r')\frac{1}{2} }-
e^{i\b l\cdot(\b r+\b r')\frac{1}{2} }+e^{-i\b l\cdot(\b r-\b r')\frac{1}{2} }\bigg]
\nonumber\\
&&\times
\bigg[ e^{-i\b l\cdot(\b r-\b r')\frac{1}{2}}-e^{i\b l\cdot(\b r+\b r')\frac{1}{2} }-
e^{-i\b l\cdot(\b r+\b r')\frac{1}{2} }+e^{i\b l\cdot(\b r-\b r')\frac{1}{2} }\bigg]\,.\label{phases1}
\end{eqnarray}
Each term in the $\b l$-integral is logarithmically divergent and requires regularization at some scale $l_m$. This scale cancels out from the final result that reads
\beq\label{phases2}
\int d^2B\, \im [i\Gamma^{dd}]=4\pi\alpha^2\bigg( r^2\ln \frac{|\b r-\b r'||\b r+\b r'|}{r^2}+r'^2\ln \frac{|\b r-\b r'||\b r+\b r'|}{r'^2}+2\b r\cdot \b r'\,\ln \frac{|\b r+\b r'|}{|\b r-\b r'|}\bigg)\,. 
\eeq

Integration over angle $\varphi$ between the directions of vector $\b r$ and $\b r'$ can be done using the integrals
\begin{eqnarray}
\int_0^{\pi}d\varphi \ln  \frac{|\b r-\b r'||\b r+\b r'|}{r^2}&=& -2\pi \ln \frac{\min\{r',r\}}{r'}\,,\label{int01}\\
 \int_0^{\pi}d\varphi\, rr' \cos\varphi  \ln \frac{|\b r+\b r'|}{|\b r-\b r'|}&=& \pi \min
\{r'^2,r^2\}\,,  \label{int02}
\end{eqnarray}
After some simple algebra  the cross section \eq{gammap} takes form
\beql{xsecU}
\sigma_\mathrm{tot}^{\gamma\gamma}(s)= \frac{16\alpha^4}{\pi m^2}
\int  \frac{d^2u}{2\pi} \, u^4 F(u)\int _0^1d\xi\, \xi ^3 F(u\xi)\ln \frac{e}{\xi}\,,
\eeq
where $\b u=m\b r$ and $\xi = r'/r$. Substituting the following Fourier images 
\begin{eqnarray}
u^2K_0^2(u)&=& \int \frac{d^2p}{2\pi}e^{-i\b u\cdot \b p}\, 
\frac{8[p(1+p^2)\sqrt{4+p^2}-(4+2p^2+p^4)\,\mathrm{arcsinh}(p/2)]}{p^3(4+p^2)^{5/2}}\,,\\
u^2K_1^2(u)&=&  \int \frac{d^2p}{2\pi}e^{-i\b u\cdot \b p}\, 
\frac{4[p(p^2-2)\sqrt{4+p^2}+8(1+p^2)\,\mathrm{arcsinh}(p/2)]}{p^3(4+p^2)^{5/2}}\,,
\end{eqnarray}
into \eq{xsecU} converts the $\b u$ integral into the delta function. Removing it, integrating over $\xi$ and then over $\b p$ gives after rather lengthy calculations the final result \eq{gg-fino}.


\end{document}